\documentclass[a4paper,11pt]{article}
\usepackage{pos}

\title{Outreach Modules for New Particle Searches Using the ATLAS Forward Proton Detector and for Higgs Boson Physics}

\author*[a,b]{Andr\'e Sopczak}
\affiliation[a]{Czech Technical University in Prague, Institute of Experimental and Applied Physics,\\
Husova 240/5, CZ-110\,00, Prague 1, Czechia}
\affiliation[b]{on behalf of the International Particle Physics Outreach Group}

\emailAdd{andre.sopczak@cern.ch}

\abstract{
We present two modules as part of the Czech Particle Physics Project (CPPP). These modules are intended as learning tools in masterclasses aimed at high-school students (aged 15 to 18). The first module is dedicated to the detection of an Axion-Like-Particle (ALP) using the ATLAS Forward Proton (AFP) detector. The second module focuses on the reconstruction of the Higgs boson mass using the Higgs boson golden channel with four leptons in the final state. The modules are accessible at http://cern.ch/cppp.}

\FullConference{%
  41st International Conference on High Energy physics - ICHEP2022\\
  6-13 July, 2022\\
  Bologna, Italy
}

\begin{document}
\maketitle

\section{Introduction}

 Outreach is an important part of CERN's mission. An audience that is particularly important consists of high-school students who might be interested in pursuing further studies in the
 Science, Technology, Engineering and Mathematics (STEM) fields. 
 A set of online materials is presented that can be used in 
 masterclasses to teach high-school students (15 to 18 years old) about particle physics at CERN. 
 These tools are part of the wider Czech Particle Physics Project (CPPP). 
  
    Two modules have been developed, each having two main components: 
    an introduction which explains the basic physics involved and 
    a simulation which allows users
    to learn about physics research interactively.
    The first module takes students through the process of finding an Axion-Like Particle (ALP) using the ATLAS Forward Proton (AFP) detector. The second shows students how to determine the mass of 
    the Higgs boson by studying the so-called golden channel 
    ($H \to ZZ \to 4e$).
Details of the implementation of the modules are given in Refs.~\cite{Zacik:2774895,ippog2021}.

\section{Czech Particle Physics Project}

The Czech Particle Physics Project (CPPP) 
is planned as a 
wide-ranging outreach project with a home page which lets the user easily access different modules.
Currently, it contains the ALP and Higgs boson golden channel modules, as well as a module that allows users 
to search a database of publications with automated update and categorization
related to the experimental Higgs boson 
research~\cite{Higgs_these_proc}\footnote{Following a feasibility study by Martin Kupka~\cite{Kupka:2722144}, implemented by Peter Zacik~\cite{Zacik:2774895}.}.
The CPPP is expected to be expanded with new modules 
added to the home page in the structure
that is provided.

\section{Finding an ALP using the AFP detector}

\subsection{ALP and AFP}

Axion-Like-Particles (ALP) are potential dark matter candidates that are predicted by some extensions of the Standard Model~\cite{Baldenegro:2729326}. 
These new particles could be detected with the ATLAS detector at CERN.
The ATLAS Forward Proton (AFP) detector is used to enhance
the sensitivity for a discovery.
The AFP detector is located on either side of the ATLAS main detector 
and it is able to detect protons that come from the central $pp$ interaction point but are deviated from the axis of the beampipe. 
Two incoming protons could pass closely by each other 
instead of strongly interacting 
in which case they will interact 
through their electromagnetic fields and be deflected. 
This can produce a pair of photons. If ALPs exist the photons could scatter through an ALP ($\gamma \gamma \to ALP \to \gamma \gamma$), such that the scattered photons are detected in the central detector while the deflected protons are detected in the AFP detector.
There are also  diphoton events without an ALP being involved 
which are referred to as  background events and 
can be largely removed by the two calculated energy loss values from 
the diphoton system and the deflected protons. 
If the scattering happened through an ALP the calculated 
energy loss should match. 
This criterion allows us to increase the ratio of 
signal-to-background events and thus isolate the signal better. 
The fact that the information from an additional detector 
enhances the sensitivity to find a new particle 
is an important learning aspect.


\subsection{Introduction page}

Before obtaining access to the simulation, 
the user reads through an introduction page which has five sections.
This introductory page contains all the information the user needs to appreciate the simulation. It should also stimulate further reading about CERN, LHC and the ATLAS detector.

\subsection{User experience}

The main components of the interactive event display are:

\begin{enumerate}
    \item Control panel: allows the user to interact with the simulation
    \item Event counters: counter for the total number of events fired and counter for auto-fired events
    \item Inner ATLAS central detector
    \item ATLAS central detector side view with AFP detectors on either side: animation of the protons being fired
    \item Energy loss matching histograms: the energy loss calculated from the AFP detectors in red and from the ATLAS central detector in green
    \item Invariant mass histograms: left histogram for all events, right histogram only for events where the energy loss values are matched (as decided by the user, or automatically in auto-fire mode)
    \item Access to admin page: password protected
\end{enumerate}

Once the user loads the simulation page, a succession of post-it notes on the screen will guide through the steps needed to run the simulation.

After going through the analysis process manually at least five times, the user unlocks the auto-fire feature. This allows the user to generate a large number of events automatically. In general, around 100 events are needed to observe a clear signal in the simulated data. 

Once the user has gone through the full auto-fire mode, a quiz pops-up which asks the user to locate the signal. If this is done correctly an explanation is given and the user is free to continue with the simulation. If a wrong answer is given, the user is encouraged to continue firing more events and collect more statistics.

\subsection{Admin page}

The website has a password protected admin page. This page allows an administrator to modify the parameters of the simulation. 
Changes on this page are sent to the server and are applied globally, 
thus anyone on the website will have the new parameters.
A possible extension of the project could be that individual user groups can define their own Admin settings.

\section{Higgs boson module}


The Higgs boson golden channel is one of the research modes of the Higgs boson~\cite{Aaboud:2277731}. 
In this channel, the Higgs boson decays into a pair of $Z$ bosons (real or virtual). Each $Z$ boson then decays into a pair of light leptons (electron or muon).
The signal for the Higgs boson is four leptons 
with an invariant mass equal to the Higgs boson mass. 
The main background to this signal is the production of di-$Z$
bosons directly from the proton collision decaying to leptons.
For simplicity, the module focuses only on the four-electron final state.

This process has been observed with the ATLAS detector~\cite{Aaboud:2277731}. 
The signal peaks at the Higgs boson mass (125 GeV). 
The background covers a wide range of invariant masses and 
peaks at the $Z$ boson mass around 91~GeV.
In this module, the user learns how to identify 
the four leptons by applying a selection 
on the lepton transverse momentum, 
and thus the user will reproduce the ATLAS distribution 
with the $Z$ and Higgs boson mass peaks.

\begin{figure}[h]
        \centering
    \includegraphics[width=\textwidth]{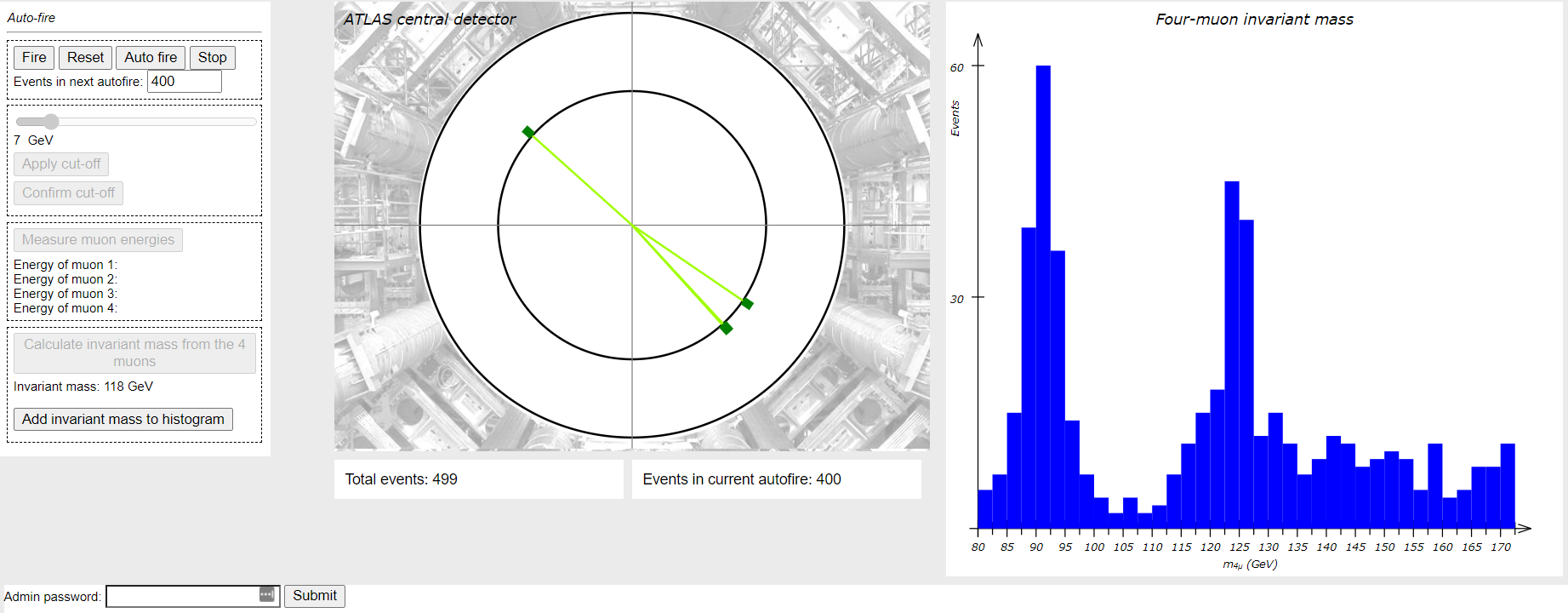}
        \caption{Screenshot of the interactive event display in the Higgs boson golden channel module.}
        \label{fig:Higgs_screenshot}
\end{figure}

The introduction page, user experience and admin page are similarly structured
as in the ALP with AFP module.
The interactive event display for this module is shown in Figure~\ref{fig:Higgs_screenshot}. 

\section{Conclusions} \label{sec:conclusions}

Two new outreach modules have been developed and implemented for the Czech Particle Physics Project.
They are accessible at \href{http://cppp.web.cern.ch}{http://cern.ch/cppp},
and are ready to serve students, teachers, and the interested public.
Learning goals are:
giving school students 
a sense of the fun with science, technology, engineering and mathematics,
awakening interest in CERN, LHC and ATLAS physics,
understanding basic concepts in particle physics,
acquiring initial knowledge about elementary particles and how to detect them,
becoming aware that additional detectors help to separate signal and background events,
improving the understanding that increasing data statistics leads to new observations and discoveries.

Feedback on the developed modules is very welcome and 
will be used in future updates of the modules. 
Further modules on related fields of research as well as 
modules on other research directions can be efficiently 
created and maintained using the structural setup in this project.

\clearpage
\section{Acknowledgements}
The author would like to 
thank the 
International Particle Physics Outreach Group
(IPPOG) for fruitful discussions, and the students
Martin Kupka, Antoine Vauterin and Peter Zacik
for their dedication.
The grant support from the "Fondu celoškolských aktivit pro rok 2022" Czech Technical University in Prague is gratefully acknowledged.
The project is also supported by the Ministry of
Education, Youth and Sports of the Czech
Republic under the project number
LTT 17018.

\bibliographystyle{JHEP}
\bibliography{biblio}


\end{document}